\documentclass[journal]{vgtc}                     

\onlineid{0}

\usepackage{xcolor}

\vgtccategory{Representations \& Interaction}

\title{Loom: Multi-Region Analysis of Spatial Transcriptomics with Local Neighborhoods and Global Trajectories}

\author{%
  \authororcid{Siyuan Zhao}{0009-0006-9834-7839},
  \authororcid{Nafiul Nipu}{0009-0006-4602-0359},
  \authororcid{Hossein Fathollahian}{0009-0001-1171-6886} \\
  Hao Chen,
  Ameen Salahudeen,
  Olga Karginova, and 
  \authororcid{G. Elisabeta Marai}{0000-0002-7212-9669}
}

\authorfooter{
  \item
      Siyuan Zhao, Nafiul Nipu, Hossein Fathollahian, Hao Chen, Ameen Salah
      udeen, Olga Karginova, and G. Elisabeta Marai are at the University of 
      Illinois at Chicago. E-mail: \{szhao69, mnipu2, hfatho2, chenhao, ameen, okar\}@uic.edu
}

\abstract{
  We present Loom, a spatial transcriptomics (ST) visual computing system to support the analysis of pseudo-temporal trajectories, comparative investigation across samples and regions of interest, and the examination of spatially structured processes within local microenvironments. ST is a molecular profiling technology that measures gene expression directly within a thin tissue section while preserving its spatial organization. For practical application-driven analyses, the ST local microenvironment data needs to be integrated with cell reference datasets and temporal simulations of cell behavior. This integration is challenging due to multi-modal registration issues and the complexity of the pseudo-temporal patterns, spatial enrichment data, and gene expression dynamics. Loom leverages a novel glyph coupled with a computational backbone to facilitate the detailed pseudo-temporal exploration of local microenvironments, cross-sample comparisons, and investigation of spatiotemporal biological mechanisms. We evaluate Loom quantitatively through a performance study, through two case studies developed with experts in tissue pathology and oncologists,  and through an external usability study. The results demonstrate that Loom supports effectively the discovery of cellular transitions and spatiotemporal expression dynamics.
}

\keywords{Life Sciences, Health, Medicine, Biology, Bioinformatics, Genomics, Mixed Initiative Human-Machine Analysis, Visual Representation Design, Application Motivated Visualization}

\teaser{
  \centering
  \includegraphics[width=\linewidth, alt={Teaser}]{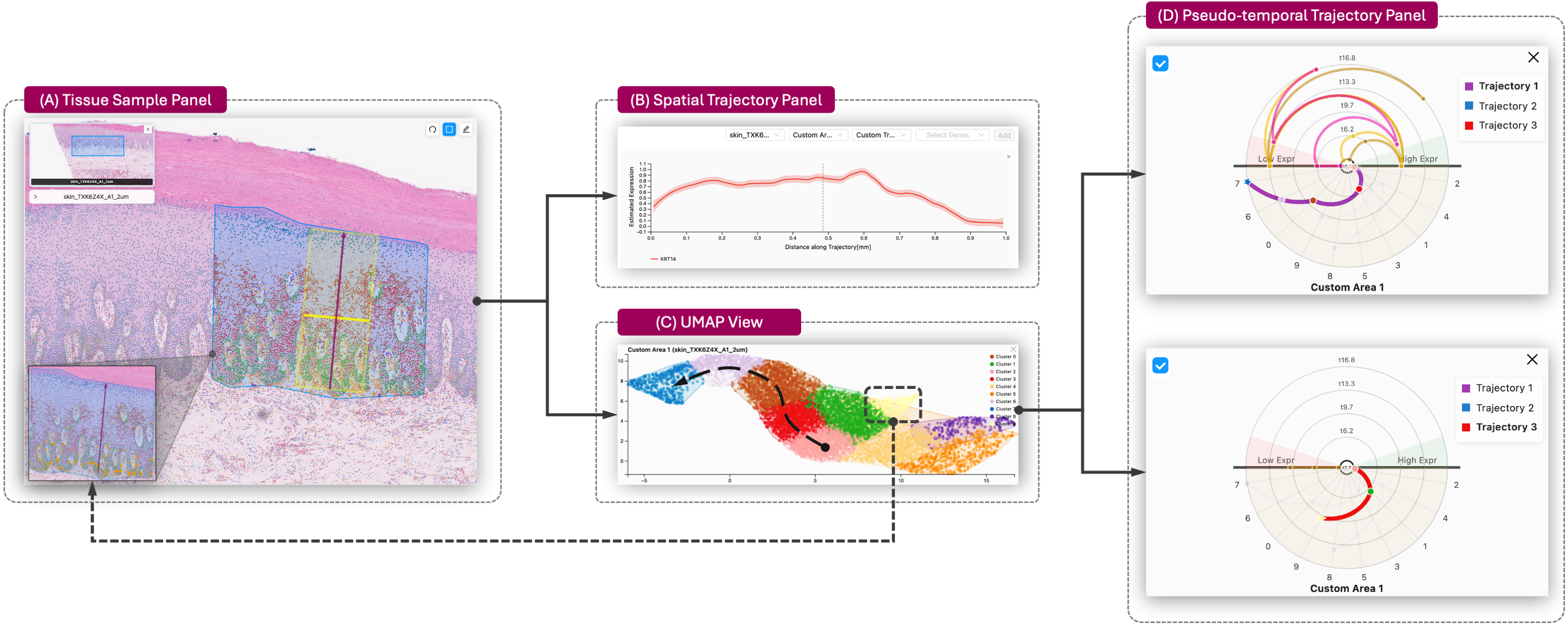}
  \caption{
  	Loom Spatiotemporal Analysis Case 1. (A) Tissue sample view showing a verified melanocyte distribution at the tissue boundary (orange dots in bottom left swatch). (B) Spatial trajectory analysis result along the selected direction (purple) and within the selected range (yellow) of the tissue sample view. (C) \textcolor{black}{UMAP of region clusters along the same purple direction for the small melanocyte population.} (D) Pseudotime analysis along the selected direction. The glyph maps time to its radius, growing outward, and expression levels horizontally. The clockwise ordering of the lower half shows the cluster assignment results of the spatial enrichment analysis. Top: dynamic expression of TOP2A (\textcolor{black}{Ochre}), KRT14 (\textcolor{black}{Amber}), and KRT10 (\textcolor{black}{Pink}), capturing early, intermediate, and differentiation-associated states. Bottom: A secondary branch toward the melanocyte population, where MITF does not show a clear temporal trend.
  }
  \label{fig:Teaser}
}

\graphicspath{{figs/}{figures/}{pictures/}{images/}{./}} 

\usepackage{tabu}                      
\usepackage{booktabs}                  
\usepackage{lipsum}                    
\usepackage{mwe}                       
\usepackage{ccicons}                   
\usepackage{amsmath}
\usepackage{amsfonts}
\usepackage{mathptmx}                  

\begin{document}


\firstsection{Introduction}

\maketitle

Spatial transcriptomics (ST) comprises a set of methods that capture the spatial context, within intact tissue, of the process of gene expression, i.e., a cell making RNA copies of DNA sequences. ST can capture local microenvironments, which play a critical role in defining the context-dependent behavior, gene expression, and functional state of cells. This spatial information is essential: for example, the behavior of immune cells varies with their position within a tumor, and the developmental states of cells are shaped by their local microenvironment. Conversely, without the spatial and local microenvironment context, biology research is limited to a “gene list” (“a fruit salad”~\cite{keller2025state}), unable to infer how gene activity organizes into biological structures or governs tissue-level functions (“a fruit tart”~\cite{keller2025state}). In contrast, in spatially resolved methods like ST, one can observe not only individual entities but also how they are organized and their spatial relationships. 

State-of-the-art ST technologies like \textcolor{black}{Visium HD} can now provide results at subcellular levels of resolution (2 µm). However, for application-driven analysis such as identification of promising drug targets, these ST images (the “fruit tart” images) need to be integrated with -omics experimental datasets (i.e., the “fruit” experimental characteristics). The ST local microenvironments also need to be integrated with temporal simulations of cell behaviors (the “fruit” and “fruit tart” evolution), because cells are, fundamentally, dynamic entities that respond over time to internal and external stimuli. To reconstruct the cell temporal behavior, biologists profile cells from separate organisms at various timepoint snapshots, then algorithmically reconstruct cellular dynamics from these snapshots. By integrating the ST local microenvironments with these temporal simulations and with additional -omics experimental data, ST methods could help provide insights in the fields of oncology, immunology, neuroscience, pathology, and histology. 

This integration for ST analysis is subject to significant challenges related to comparing spatiotemporal variations in gene expression across multiple samples and regions of interest,  in particular multi-modal registration and transforming and visualizing these complementary, complex, spatiotemporal data. Temporal progression is also inferred rather than directly observed, and thus requires verification. Due to these challenges, current ST platforms offer limited analysis workflows that combine 2D tissue images, overlaid gene expression heatmaps, static cluster labels, and dimensionality reduction or neighborhood-based spatial statistics. Temporal cell simulations are not leveraged. As a result, biologists struggle to trace clustering patterns or the dynamic drivers of observed differences in gene or cell distributions.

In this work, we introduce a visual analysis system to integrate temporal simulations with global spatial structures and local tissue microenvironments, and to support the interactive comparison of multiple ST datasets or regions. Our first contribution is a pre-processing and computational backend that leverages a spatial coordinate registration algorithm along with dimensionality reduction, clustering, and pseudotime simulation. Our second contribution is the design of a novel, compact visual encoding, inspired by natural plant growth, to encode pseudo-temporal patterns, spatial enrichment, and gene expression dynamics. The encoding is leveraged in our third contribution, a novel visual computing environment, Loom, which integrates a computational backbone with coordinated multi-view visualizations in order to support ST spatiotemporal analysis in the context of local microenvironments. Our fourth contribution is a validation of Loom through a performance study, two case studies developed with domain experts, \textcolor{black}{and a quantitative usability evaluation with external participants}.

\section{Background and Related Work}

\subsection{Background}
Genes are small segments of DNA, a nucleic acid essential for life, located within a cell that provide instructions for specific functions. DNA primarily serves as the long-term storage of genetic information, while RNA, a second type of nucleic acid, acts as a messenger and plays a crucial role in protein synthesis. Transcriptomes are the sum total of all the messenger RNA molecules expressed (i.e., used to produce a functional product) from the genes of an organism, and are indicative of gene activity. Historical biological analysis methods, known as single-cell analysis, use tissue dissection followed by cell and gene sequence analysis to understand the mechanisms of certain diseases.  
 
\textcolor{black}{Visium HD}, a state-of-the-art high-resolution ST technology, can currently attain subcellular levels (2 µm). To reconstruct single-cell locations from the \textcolor{black}{Visium HD} gene expression data, state-of-the-art approaches like  bin2cell~\cite{polanski2024bin2cell} use a combination of deep learning-based cell segmentation models to identify the actual locations and boundaries of nuclei within the tissue image, and spatial clustering of gene expression to help define cell boundaries. Meanwhile, an increasing number of large, well-annotated single-cell -omics atlases have become available, providing rich reference information on cell types. 

Last, cells are also dynamic entities that respond over time to internal and external stimuli. As many experimental methods result in destruction of the cells, biologists use workarounds, where they profile cells from separate organisms at various time-point snapshots. Several approaches have been designed for reconstructing cellular dynamics, including algorithmic techniques which enable inference of dynamics from standard single-cell -omics experiments that would otherwise be viewed as snapshots. For example, pseudotime algorithms assume that similar cells in high-dimensional space are closer developmentally and can predict cell orderings. Additional tools like STLearn~\cite{pham2023robust}, GASTON~\cite{chitra2025mapping}, and SPATA2~\cite{kueckelhaus2024inferring, hao2024dictionary} can model spatial trajectories.

\subsection{Related Work}
\textbf{Spatial Transcriptomics Visualization.} Several research toolkits~\cite{palla2022squidpy, wolf2018scanpy, dries2021giotto, sztanka2022spacemake, schapiro2017histocat, chen2025giotto} and commercial platforms~\cite{10xgenomics, cellxgene, bioturing} provide analytical methods for exploring complex ST datasets. Commercial platforms provide users with convenient access to basic exploration tasks, such as spatial mapping and cell type inspection. Research toolkits, often implemented as Python or R packages, expose a wider range of computational methods for tasks such as nuclei segmentation and neighborhood, etc., yet rely on external visualization components for result interpretation. To address these limitations, several visualization systems~\cite{fan2020spatialdb, keller2025vitessce, fernandez2019st, fathollahian2025attention, li2023spacewalker, li2025genesurfer, zhou2024sorc, yang2024srt, ospina2025spatialge, fan2023spascer, zheng2023aquila, xu2024stomicsdb, li2025soar, yuan2023sodb, deng2024scar} seek to integrate pre-computed analysis outputs with interactive visual exploration. For example, Vitessce supports flexible composition of derived data within coordinated multi-view layouts, although it does not support re-clustering. None of these tools supports pseudo-temporal analysis.

With respect to comparative ST visualization, several software platforms provide ST region-selection mechanisms~\cite{honcharuk2025deepspacedb, solorzano2020tissuumaps, bankhead2017qupath, lewis2021spatial, cao2023smdb, bergenstraahle2020seamless, warchol2025seal} for comparison. However, the analytical support provided is limited to aggregated statistics. These systems also overlook intra-regional directional patterns (e.g., gradient variations) in the ST data. Specifically, these directional processes are inherently anisotropic, with variations unfolding unevenly across space and often following specific directions, such as tumor invasion fronts or localized immune responses. As a result, statistical analysis of regions as homogeneous units can obscure meaningful directional patterns. Existing software platforms do not support an analytical approach that can both preserve spatial context and enable fine-grained, directionally sensitive analysis.

Overall, these existing ST visualization systems support spatial inspection, clustering-based exploration, and basic multi-sample visualization. They typically treat computational analysis and visualization as separate stages, with advanced methods such as pseudo-temporal inference or spatial gradient analysis performed externally and incorporated only as static results. Multi-sample functionality is also typically restricted to parallel inspection or comparison of aggregated outputs, rather than enabling coordinated, region-level analysis across samples. Moreover, existing systems provide limited support for analyzing spatially structured processes within user-defined regions, especially those involving directional variation or continuous transitions. In contrast, Loom integrates these analytical components within a unified visual computing system, enabling interactive, region-based spatiotemporal analysis and coordinated comparison across samples.

\noindent \textbf{Spatial-temporal and Temporal Visualization.} Spatio-temporal visualization research explores ways to encode both space and time, including space-time cubes~\cite{kraak2003spacetime, bach2014review}, flow or vector field visualizations~\cite{nipu2023visual, nipu2026neurosurgery, luciani2018details}, dynamic network layouts~\cite{boyandin2011flowstrates}, and biological imaging tools to track cellular motions\cite{amat2014fast, wolff2018mamut, aurisano2015bactogenie, luciani2014fixingtim, FATHOLLAHIAN2026104728, chen2026chrompolymerdb}. However, these general methods ~\cite{longo2021integrating, moses2022museum,torkamani2015video} do not seek to capture spatial organization, as the one in ST. 

Many visualization techniques address temporal data without spatial information, including glyph-based encodings~\cite{fischer2012clockmap, thakur2009data, tang2019treeroses}, dendritic or tree-like representations~\cite{lin2007experiencing, burch2008timeline}, and circular or pie-chart–based encodings~\cite{krzywinski2009circos, Sheidin2017TimeRay,fathollahian2025peaks}. Domain-specific visualization methods have also been developed to address challenges in analyzing time-series data~\cite{Aigner23TimeVizSecondEdition}, such as Filipov et al.'s circular technique for representing multiple related historical events~\cite{filipov2021gone}. Current ST workflows use similar encodings to separate temporal changes for cell clusters and spatial interactions across clusters~\cite{cao2019single, jin2021inference}. In contrast, we seek to simultaneously encode both spatial context and temporal progression.

\section{Methods}

\subsection{Requirements Analysis}
This project was developed as a collaborative research project between researchers in the University of Illinois College of Medicine and visual computing researchers at the University of Illinois Chicago. In designing our visual analysis solution, we adopted an Activity-Centered Design (ACD) strategy~\cite{marai2017activity}. Following this perspective, our interdisciplinary team iteratively mapped key analysis routines, created parallel low- and high-fidelity prototypes, and refined visual encodings and interactions through multiple evaluation cycles. Previous design-study research has shown that ACD tends to perform particularly well in complex, cross-domain collaborations, achieving substantially higher success rates than traditional Human-Centered Design approaches~\cite{marai2017activity}. Requirements for both the interface and workflow were initially gathered through interviews with three research oncologists and subsequently refined during meetings, followed by multiple rounds of prototyping and feedback over several months. Feedback from these collaborators was incorporated into the design.

\subsection{Data Analysis}
Our collaborators' workflow starts with profiling tissue samples using Visium HD. The project ST data consists of \textcolor{black}{Visium HD} human tissue samples. Each sample was measured at three spatial resolutions (2 µm, 8 µm, and 16 µm). On average, each sample included over 100,000 cells with gene expression profiles covering approximately 18,000 genes per cell. Although \textcolor{black}{Visium HD} data do not represent true single cells, the resolution is sufficiently close to the single-cell level. For clarity, we therefore use the term “cell” rather than “spot” throughout this paper. The result is paired with high-resolution histological images on the order of several billions of pixels, resulting in a total data volume of roughly 150 GB per sample. These measurements form global, large-scale heterogeneous multivariate spatial data that integrate high-dimensional gene expression and multi-resolution spatial image data. 

Raw measurement results are often noisy and not directly comparable across cells. Consequently, our collaborators follow a multi-stage workflow that progressively transforms the data through preprocessing, such as quality filtering to remove low-quality measurements and normalization to ensure comparability, into a sequence of derived analytical representations~\cite{heumos2023best}. To reveal the latent structure in high-dimensional data, this workflow first employs UMAP projection, which generates a derived dataset consisting of 2D embedding coordinates. This reduction process organizes multivariate data into distinct sets or clusters based on the similarity of multidimensional features, effectively translating complex mathematical correlations into spatial proximity. Consequently, these derived coordinates could serve as a critical data layer for visualization; we could map this structural information directly to the spatial layout, where the resulting clusters represent groups of entities sharing similar evolutionary behaviors and parametric properties.

Next, for pseudo-temporal analysis, our collaborators model the evolution of cell states over time using Slingshot, a state-of-the-art computational tool for analyzing single-cell transcriptomics data to infer cell lineage and progression through developmental stages, as detailed in the work by Street et al.~\cite{street2018slingshot}. These temporal trajectories are typically visualized as color-gradient scatter plots, enabling biologists to track dynamic changes in cell states. However, this representation primarily emphasizes the overall trends of change along an inferred temporal axis. Furthermore, cellular tissue positions, neighborhood structures, and regional boundaries are typically not directly depicted in this representation. Consequently, our collaborators faced difficulties correlating branching, fluctuations, or state transitions within the trajectory with the spatial organization of the original tissue.

Last but not least, our collaborators are interested in the local spatial information. Local neighborhoods, defined by the spatial proximity of mapped regions, allow the quantification of interactions and microenvironmental effects among neighboring cells. For this last workflow step, our collaborators use spatial neighborhood enrichment analysis~\cite{dries2021giotto, palla2022squidpy} to identify statistically significant co-localization patterns and interaction tendencies among cell populations; however, this analysis primarily captures static co-occurrence relationships and does not explicitly incorporate dynamic progression.

\subsection{Task analysis}
Through discussions with our collaborators, who are experts in tissue pathology and oncologists, we learned that many critical questions arise not from analyzing a single sample in isolation, but from comparing patterns across multiple conditions, such as different disease stages, distinct disease subtypes, or contrasts between healthy and diseased tissues. These comparisons require integrating heterogeneous information, including tissue morphology, local microenvironmental context, and sample-specific clinical findings. They also emphasized that the spatial organization of cells and the structure of local neighborhoods often carry more biologically interpretable information than standard clustering or differential expression results, providing insights into how tissues maintain function or respond to pathological processes. They expressed frustration related to the fact that their typical workflows rely on a collection of disparate analysis tools, most implemented in separate environments.

As a result, the Loom top-level design focuses on supporting multi-sample comparison, region-based comparison, and spatiotemporal analysis, enabling researchers to quickly identify differences across biological conditions, examine how cellular relationships and tissue organization vary across space and along inferred temporal progression, and explore hypotheses about the underlying biological mechanisms driving these differences. Based on interviews and feedback from the prototyping stage, we summarized the following activities and task abstraction (GE denotes gene expression):
\newcommand{\analysisheading}[1]{%
    \par\addvspace{0.5\baselineskip}%
    \noindent\textbf{#1}\par\nobreak
}

\analysisheading{A1. Analysis of spatial organization within tissue samples}

\begin{enumerate}[ label=T1.\arabic*, leftmargin=2.5em, topsep=0.4em, partopsep=0pt, parsep=0pt ]
    \item Delineate regions of interest (e.g., neighborhoods)
    \item Detect spatially proximal cell and neighborhood structures
    \item Characterize neighborhoods via cell types and GE
    \item Identify GE co-localization or gradient patterns
    \item Examine GE changes along the spatial trajectory
\end{enumerate}

\analysisheading{A2. Comparative spatiotemporal analysis across samples and ROIs}

\begin{enumerate}[ label=T2.\arabic*, leftmargin=2.5em, topsep=0.4em, partopsep=0pt, parsep=0pt ]
    \item Compare GE spatial patterns along pseudo-temporal trajectories across samples
    \item Identify region-specific differences in progression dynamics
    \item Analyze cell compositions along spatial trajectories under different conditions
    \item Detect regions with distinct temporal progression behaviors
    \item Relate spatiotemporal variations to underlying biological processes or pathways
\end{enumerate}

\section{Design}

\subsection{Architecture}

\begin{figure*}[htb]
  \centering
  \includegraphics[width=0.9\textwidth]{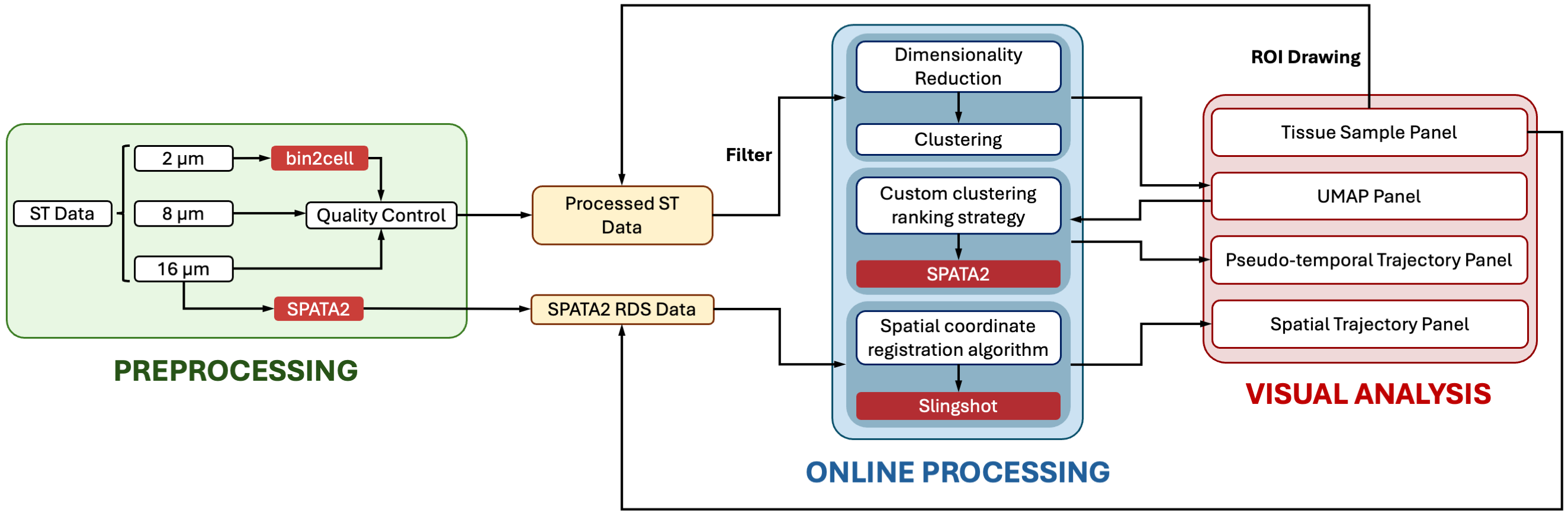}
  \caption{
    \label{fig:Architecture}
    Loom architecture: offline preprocessing component, online processing backend, and interactive visual analysis. 
    }
\end{figure*}

The Loom overall architecture (Fig.~\ref{fig:Architecture}) has four core computational modules: (1) preprocessing of ST data, (2) dimensionality reduction and clustering for ROI-specific analyses, (3) a spatial coordinate registration algorithm that maps user-defined geometric features across resolutions, and (4) a clustering-based ranking strategy designed to quantify spatial interaction breadth. Together, these components form a unified backend computing system. The front-end includes custom and novel encodings for representing the data we abstracted in the section above.

The Loom computational backend is implemented in Python, with data processing handled through a Flask-based service. In addition to core libraries such as pandas, we integrate Squidpy~\cite{palla2022squidpy} for ST analysis and rpy2~\cite{haslwanter2016introduction} to interface with R-based analysis packages. The Loom frontend is built with React and D3.js~\cite{bostock2011d3}, leveraging deck.gl~\cite{wang2019deck} for GPU-accelerated rendering to support the interactive visualization of large-scale spatial data. \textcolor{black}{The source code is publicly available at \url{https://github.com/ScheWann/Loom}.}

\subsection{Preprocessing}
Performing trajectory inference directly on high-resolution ST data generates substantial noise, suppresses feature-gene detection, and increases computational cost. To address these issues, we preprocess the data as follows. We employed bin2cell~\cite{polanski2024bin2cell} to reconstruct single-cell locations, due to its ability to reproduce fine morphological structures, and we used the R-framework SPATA2 to generate spatially-resolved trajectories. The data can be influenced by biological factors, such as cells with little or no detectable gene expression. These noise signals can substantially affect downstream analysis and interpretation: excessive filtering risks eliminating rare cell subpopulations, while overly lenient filtering allows low-quality cells to remain. Following best-practice recommendations~\cite{heumos2023best}, we applied an adaptive quality control approach using the Median Absolute Deviation (MAD) method across all samples and resolutions to consistently identify and remove low-quality cells. Three recommended metrics were considered for each cell: overall gene expression, the number of detected genes, and the proportion of mitochondrial transcripts. Cells deviating by $>$5 MADs from the median were flagged as outliers and removed from the dataset~\cite{germain2020pipecomp}, where \textcolor{black}{$X_i$ is the value of a given quality-control metric for cell $i$, and $X$ is the distribution of that metric across all cells}:

\begin{equation}
  \mathrm{MAD} = \mathrm{median}\left( |X_i - \mathrm{median}(X)| \right)
\end{equation}

\textcolor{black}{Following quality control, we normalized each cell's gene expression counts to account for differences in total counts per cell, ensuring comparable expression levels across cells.} We used the shifted logarithm, which can effectively stabilize variance and reveal the underlying structure of the dataset~\cite{ahlmann2023comparison}.

\subsection{Dimensionality Reduction and Clustering}
Although highly variable genes (HVGs) are identified during preprocessing, their variability often shifts within a specific region of interest (ROI), as local tissue composition and microenvironmental heterogeneity can alter variation. To ensure that downstream analyses preserve features relevant to the current region, we re-identify HVGs within the selected ROI before conducting dimensionality reduction and clustering through the {\textcolor{black}{Scanpy’s seurat\_v3 HVG-selection method~\cite{stuart2019comprehensive}}}, which models mean-variance relationships of gene expression and identifies correspondences between cells in different samples. Given the single-cell resolution and huge number of cells and genes, even sub-regions remain high-dimensional and inherently redundant. Thus, we apply principal component analysis (PCA) to reduce noise while preserving key information. Due to the sparsity and non-linearity of single-cell data, we select the top 30 components for downstream analysis.

For subsequent dimensionality reduction, both t-distributed stochastic neighbor embedding (t-SNE) and uniform manifold approximation and projection (UMAP) are robust options. We selected UMAP based on domain expert preference. Because UMAP features remain abstract, we performed clustering to infer cell identities in the PCA-reduced space and connect each cell to its K-nearest neighbors (KNN). Dense regions in this KNN graph correspond to closely related cells in the expression space, capturing the data's topological structure. We then apply the Leiden algorithm~\cite{traag2019louvain}, an improved version of the Louvain algorithm~\cite{blondel2008fast} shown to outperform other single-cell RNA data clustering methods~\cite{weber2016comparison, duo2020systematic}, to detect clusters. \textcolor{black}{Clusters identified in the PCA-reduced space may appear continuous or partially overlapping in the UMAP visualization. Such visual overlap does not necessarily indicate that the corresponding cells are indistinguishable in the clustering space.} K was set to 15, and the Leiden resolution parameter was set to 1. These values correspond to widely adopted default choices in standard analysis pipelines, as they generally provide stable neighborhood structures and balanced cluster granularity without requiring additional tuning. Nevertheless, our system allows these values to be adjusted through the front-end.

\subsection{Spatial Coordinate Registration Algorithm}
\vspace{-1em}
\begin{figure}[htb]
  \centering
  \includegraphics[width=1.0\linewidth]{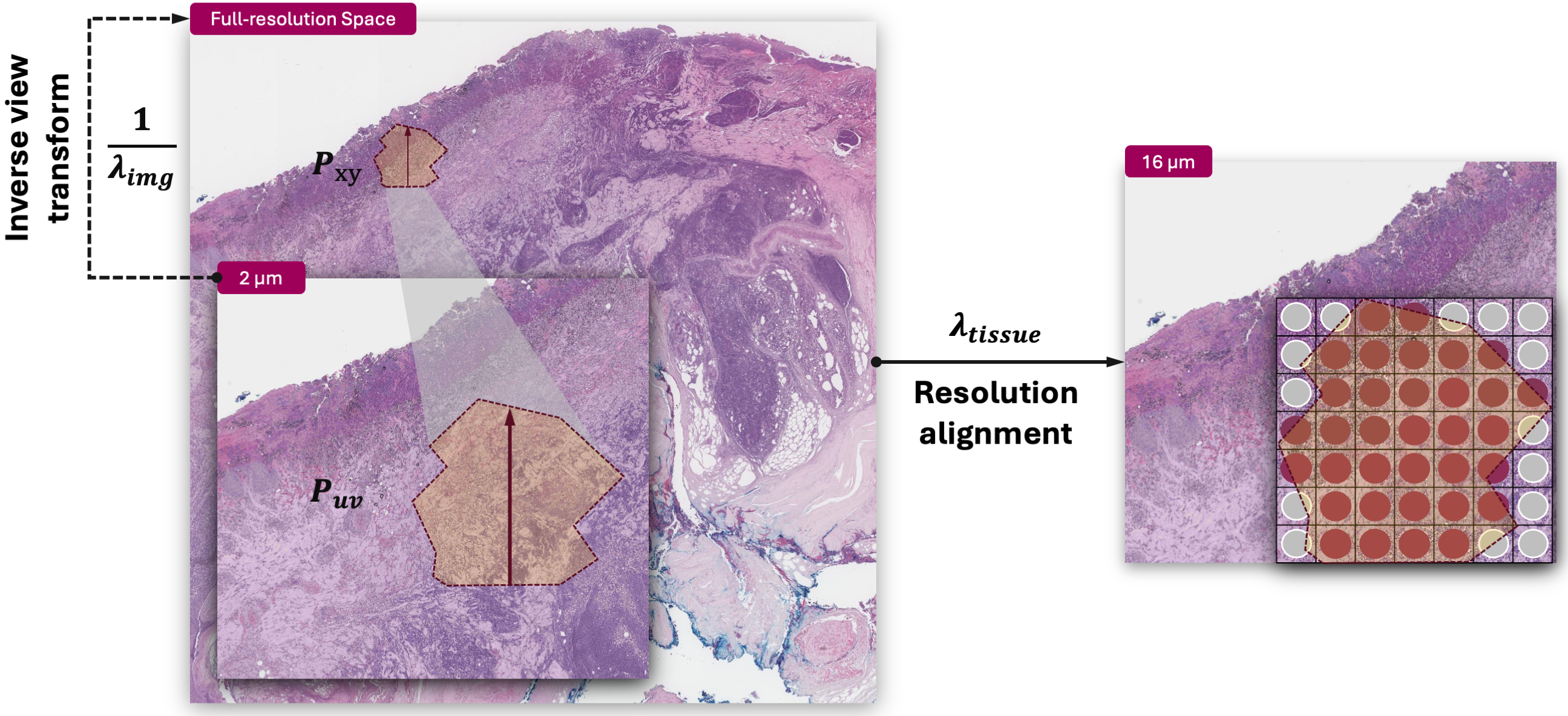}
  \caption{
    \label{fig:Spatial_coordinate_registration_algorithm}
    ROI and trajectory defined in screen space are mapped to full-resolution tissue coordinates and aligned to the 16 µm analysis space.
    }
\end{figure}

\noindent To identify spatially-coherent transcriptional patterns, we extract SPATA2 spatial trajectories (continuous paths) describing how gene expression varies across different tissue areas. After defining regions of interest (ROIs), the ROIs are mapped to the underlying spatial coordinates, the corresponding tissue regions are selected, and their expression profiles are extracted. SPATA2 then infers trajectories along these paths, identifying feature genes and generating smoothed expression trends that reflect local biological transitions. Initially, running SPATA2 on a high-resolution ROI (e.g., 8µm) introduced noise. Following expert guidance, we used the corresponding region sampled at 16 µm, which effectively captured spatial structure while minimizing noise.

Because users interact with the tissue at multiple resolutions (2 µm, 8 µm, and 16 µm), the same ROI may contain different numbers of cells/coordinates at each resolution. \textcolor{black}{Furthermore, because each resolution is cropped independently, with resolution-specific offsets and image dimensions, an ROI defined under one resolution does not align with the coordinate space of another.} To correctly align ROIs drawn at any resolution to the required 16µm resolution, we designed a spatial coordinate registration algorithm that maps user-defined geometric features (spatial trajectory vectors and ROIs) back to the original full-resolution space and then into lower-resolution bin space for analysis.

Let $P_{uv} = (u, v)$ be the user-defined screen space coordinates, and $P_{xy} = (x, y)$ the physical coordinates. First, we dynamically recover the crop offset $(x_0, y_0)$ by computing the median coordinate differences between the full-resolution matrix $S_{full}$ and the crop view matrix $S_{crop}$. We then apply an inverse affine transformation using the image scaling factor $\lambda_{img}$.

\begin{equation}
    \begin{bmatrix} x \\ y \end{bmatrix} = 
    \begin{bmatrix} x_0 \\ y_0 \end{bmatrix} + 
    \frac{1}{\lambda_{img}} \begin{bmatrix} u \\ v \end{bmatrix}
\end{equation}

To convert to the 16 µm analysis space coordinates $P'_{xy}$, required by SPATA2, we apply a down-sampling factor $\lambda_{tissue}$.

\begin{equation}
    P'_{xy} = P_{xy} \cdot \lambda_{tissue}
\end{equation}

This transformation is also applied to the width parameter $w$ of the trajectory vector to ensure geometric consistency:

\begin{equation}
    w' = \left( \frac{w}{\lambda_{img}} \right) \cdot \lambda_{tissue}
\end{equation}

Finally, to precisely define the analysis scope, we implement a geometrically constrained spatial cell selection strategy. Using Parquet-stored spatial coordinates, we construct user-defined ROI polygon boundaries $\mathcal{R}$ and use a ray-casting algorithm to determine whether each point $p_i$ lies inside. The filtering function $f(p_i)$ is defined as:

\begin{equation}
    f(p_i) = \begin{cases} 
    1 & \text{if } p_i \in \text{interior}(\mathcal{R}) \cup \partial\mathcal{R} \\
    0 & \text{otherwise}
    \end{cases}
\end{equation}

Only points with $f(p_i)=1$ are passed to SPATA2, ensuring noise-reduced and spatially accurate trajectory extraction.

As noted earlier, high-resolution ST data introduce substantial noise and computational challenges, raising concerns about whether cross-resolution analysis may compromise spatial accuracy. In the Supplemental Materials, we validate the correctness of our ROI mapping across resolutions, which underpins the use of a coarser analysis space (16 µm). The additional results on a total of \textcolor{black}{400} ROIs demonstrate that the position and shape of the ROIs are matched during the mapping process, confirming that our aggregation preserves spatial fidelity. 

\subsection{Clustering Ranking Strategy}
To integrate pseudo-temporal simulations of cell states, we develop a clustering ranking strategy. Pseudo-temporal analysis typically infers continuous trajectories of cellular state changes. However, simple pseudo-time series may overlook spatially adjacent cells and lack spatial context. To address this limitation, we integrated Neighbors Enrichment Analysis (NEA)~\cite{palla2022squidpy} with the pseudo-temporal results. NEA was selected because it extends beyond standard single-cell workflows based on clustering and differential expression, which are often limited to non-spatial data. In certain tissue contexts, clustering results alone may be difficult to interpret, whereas NEA provides spatially coherent regions that are more readily interpretable and often align with known morphological structures. By combining pseudo-time trajectories with neighboring cell information, we can identify cell populations that not only undergo state transitions along the pseudo-time axis but are also highly enriched spatially. Since NEA typically produces a 2D matrix, where each entry represents the interaction between two clusters, additional processing is needed to effectively integrate these spatial interactions with pseudo-temporal trajectories. Therefore, we implemented a clustering ranking strategy to simplify the NEA results from 2D to 1D, enabling a more straightforward interpretation alongside pseudo-temporal dynamics.

To assess the spatial “sociality” of each cell cluster, we define a “significant spatial interaction index” for each cluster $C_i$. The calculation process involves two key steps: diagonal masking and threshold counting. First, to eliminate the interference of cell cluster aggregation on connectivity assessment, we forced the diagonal elements of the Z-score matrix obtained from the spatial neighborhood enrichment analysis to zero (i.e., setting $Z_{ii} = 0$). This ensures that the evaluation index only reflects the spatial association between this cluster and other cell clusters. Subsequently, we applied a statistical significance threshold $\theta = 2.0$, corresponding to a significance level of approximately $p < 0.05$ in a standard normal distribution. For each cell cluster, we calculate the total number of interaction events $N_i$ between it and all other non-self clusters satisfying:

\begin{equation}
N_i = \sum_{j \neq i} \mathbb{I}(Z_{ij} > 2.0)
\end{equation}

where $\mathbb{I}$ is the indicator function. This index, $N_i$, effectively quantifies the spatial richness of the cell cluster's "neighborhood types."

Finally, all cell clusters are sorted in descending order based on the calculated interaction counts $N_i$. Top-ranked cell clusters tend to co-locate spatially with a diverse range of cell types, potentially playing a core role in connectivity for tissue homeostasis or signal transduction. Conversely, lower-ranked cell clusters are considered isolated populations, often forming specific regions of a single component with less spatial overlap with other cell types. This sorting strategy eliminates self-aggregation interference and, by incorporating statistical significance and focusing solely on the interaction strength of external clusters, provides highly generalizable spatial enrichment results. 

\subsection{Visual Analysis: Layout and Workflow}
Due to the complexity of the ST data and the need to compare multiple tissue regions, our top-level design for the front-end is based on a multiple coordinated view paradigm. Our layout leverages visual scaffolding~\cite{marai2015visual} from encodings which are familiar to the domain experts towards a new powerful encoding. The layout follows a streamlined workflow that begins with the selection of one or more ROIs in the tissue image, followed by detailed spatiotemporal analysis of the cell populations within those regions.

The final Loom design incorporates qualitative feedback from our collaborators. The visual frontend consists of four main panels: (i) a Tissue Sample Panel provides essential sample context and entry points for downstream analyses (T1.1, T1.2, T1.3, T1.4); (ii) a UMAP Panel supports the interpretation of biological terms within a low-dimensional embedding (T1.2, T1.3, T1.5); (iii) a Spatial Trajectory Panel reveals directional gene expression changes within selected ROIs (T1.5, T2.2, T2.3, T2.5); and (iv) a Pseudo-temporal Trajectory Panel reconstructs across cluster states and gene changes (T2.1, T2.4, T2.5). To support flexible exploration, users can run multiple clustering processes on the same ROI by adjusting parameters, and save each clustering result for subsequent spatial or pseudo-temporal trajectory analysis. At any stage, users can fine-tune the clustering results and iterate for further analyses. 

\subsubsection{Tissue Sample Panel}

\vspace{-.1em}
\begin{figure*}[htb]
  \centering
  \includegraphics[width=0.9\linewidth]{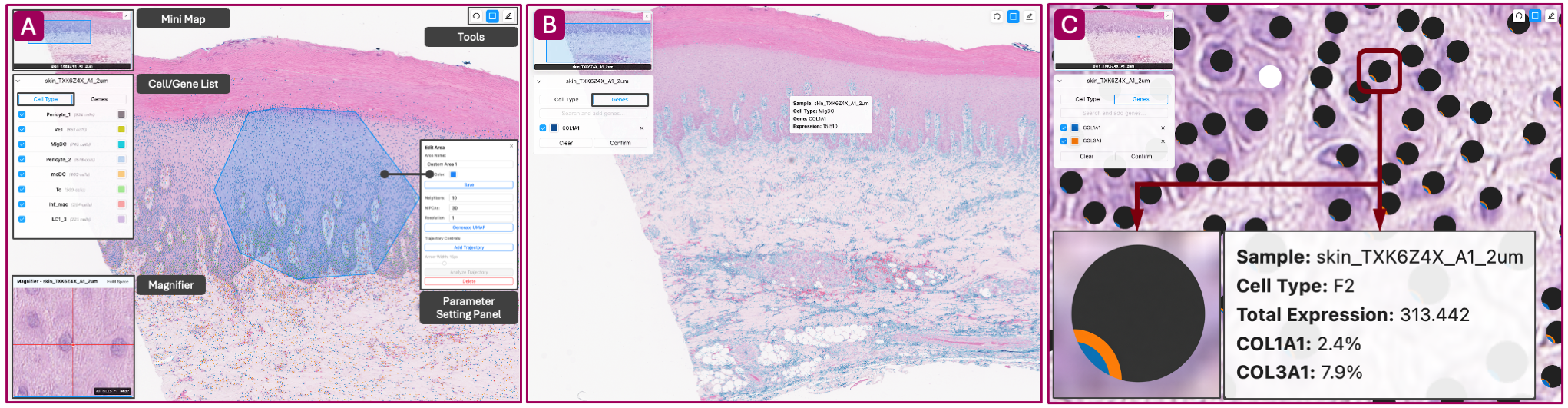}
  \caption{
    \label{fig:Tissue_Sample_Panel}
    Tissue Sample Panel. (A) Overview with a user-defined region (blue) and key components: mini-map, Cell/Gene List, magnifier, tools, and parameter settings. (B) Single-gene view showing spatial expression of COL1A1. (C) \textcolor{black}{Multi-gene view showing bin2cell-reconstructed cells at their tissue positions; Kosara’s part-to-whole encoding represents COL1A1 (blue), COL3A1 (orange), and all other genes (black) as expression proportions.}
    }
\end{figure*}

\noindent Mapping cell coordinates onto tissue images and overlaying gene or cell distributions is a fundamental function. Our Tissue Sample Panel displays multiple high-resolution histological images alongside corresponding cell distributions at different spatial resolutions. Users can zoom into regions of interest, draw and adjust arbitrary boundaries (T1.1), and use a mini-map (\textcolor{black}{Fig.}~\ref{fig:Tissue_Sample_Panel}.A) for efficient navigation and sample switching. A sample list on the left shows pre-processed cell annotations (\textcolor{black}{Fig.}~\ref{fig:Tissue_Sample_Panel}.A) (T1.2, T1.3), generated via CellTypist~\cite{xu2023automatic, dominguez2022cross}, along with a gene tab for querying and visualizing genes of interest.
In single-gene mode, expression intensity is overlaid on the tissue image (\textcolor{black}{Fig.}~\ref{fig:Tissue_Sample_Panel}.B). In multi-gene mode, we adopt Kosara’s encoding~\cite{kosara2019impact} to represent expression ratios (\textcolor{black}{Fig.}~\ref{fig:Tissue_Sample_Panel}.C), which provides a compact and effective alternative to stacked bar charts~\cite{skau2016arcs, zhao2024part} (T1.4). To mitigate occlusion from cell overlays, we incorporate a magnifier window, enabling clear inspection of underlying tissue morphology without relying on transparency adjustments.

\subsubsection{UMAP Panel}
The UMAP Panel is linked with the Tissue Sample Panel and uses a familiar scatterplot with a qualitative color scheme~\cite{harrower2003colorbrewer}. Researchers seeking fine-grained cell-type labeling may prioritize highly detailed clustering results, or, in contrast, aim to discover or validate broader biological processes, requiring iterative refinement of the clustering results. In the UMAP Panel, users can interactively adjust clustering parameters within the interface and update the results repeatedly until they obtain the level of granularity that aligns with their analytical needs. To validate a cluster’s cell type or associated biological process, we embed Gene Ontology (GO) analysis~\cite{ashburner2000gene, fang2023gseapy} to identify potential functional trends. When a cluster is selected, we automatically compute its marker genes and display the five GO terms with the highest enrichment. Cell names can be renamed once their type has been confirmed via GO analysis. The reannotated cells are then added back to the cell type list for the corresponding sample in the Sample Panel. Through iterative verification by experts, the dataset becomes progressively cleaner and more interpretable (T1.2, T1.3, T1.5).

\subsubsection{Spatial Trajectory Panel}
\begin{figure}[htb]
  \centering
  \includegraphics[width=0.9\linewidth]{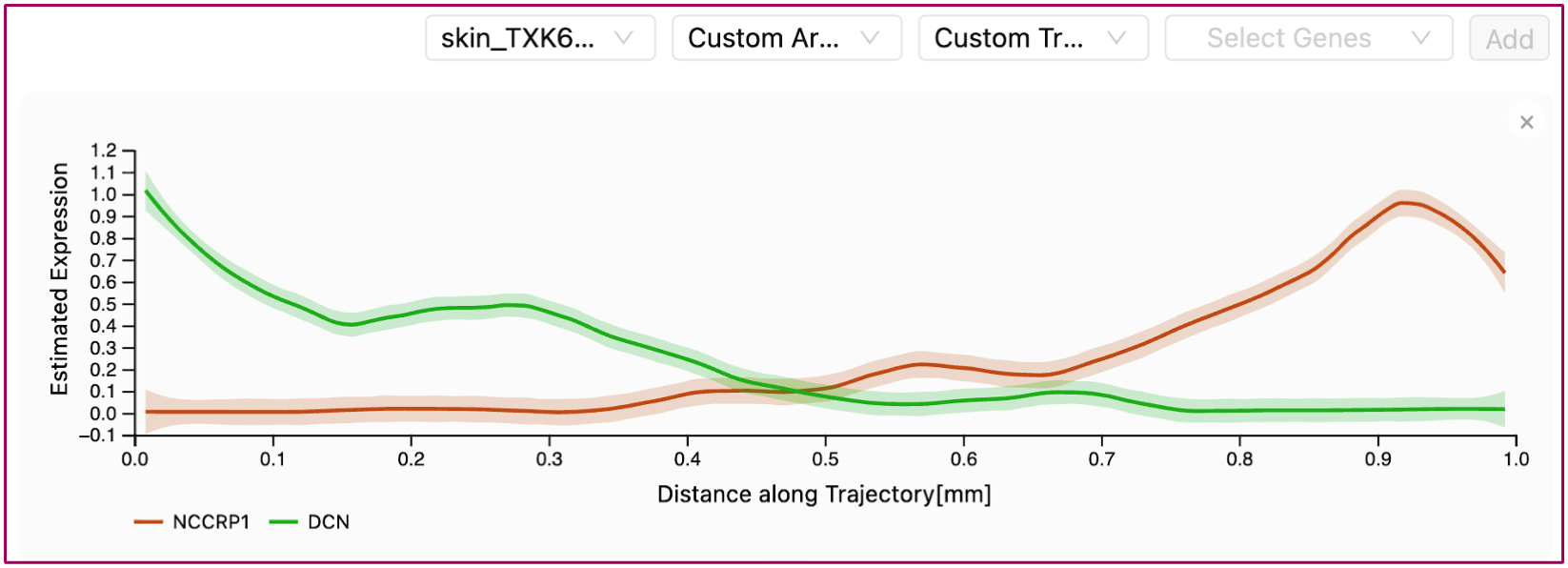}
  \caption{
    \label{fig:Spatial_Trajectory_Panel}
    Spatial Trajectory Panel: gene expression dynamics along user-defined trajectories (color curves) within a selected region, with local maximum and minimum expression levels (color bands).
    }
\end{figure}

\noindent Discussions with our cancer biology expert emphasized that disorder is a key feature of cancer cells under the microscope: cells may disperse in all directions from the lesion center or preferentially along specific directions. To address this concern, the Spatial Trajectory Panel (Fig.~\ref{fig:Spatial_Trajectory_Panel}) incorporates SPATA2 spatial trajectory analysis, which focuses on assessing whether the expression patterns of specific genes are non-randomly associated with defined spatial trajectories. For example, there are continuous changes in gene expression along the direction of tumor invasion. In this panel, after users specify an ROI and define a trajectory in the Sample Panel, the system returns a list of genes that represent candidate key genetic mechanisms. The expression dynamics of these selected genes along the plotted trajectory are visualized as line graphs (T1.5, T2.2, T2.3, T2.5). Similar to the UMAP Panel, the Spatial Trajectory Panel is linked to the Sample Panel: hovering over a trajectory automatically highlights the corresponding spatial location on the tissue image.

\subsubsection{Pseudo-temporal Trajectory Encoding and Panel}

\begin{figure*}[t]
  \centering
  \includegraphics[width=1.0\linewidth]{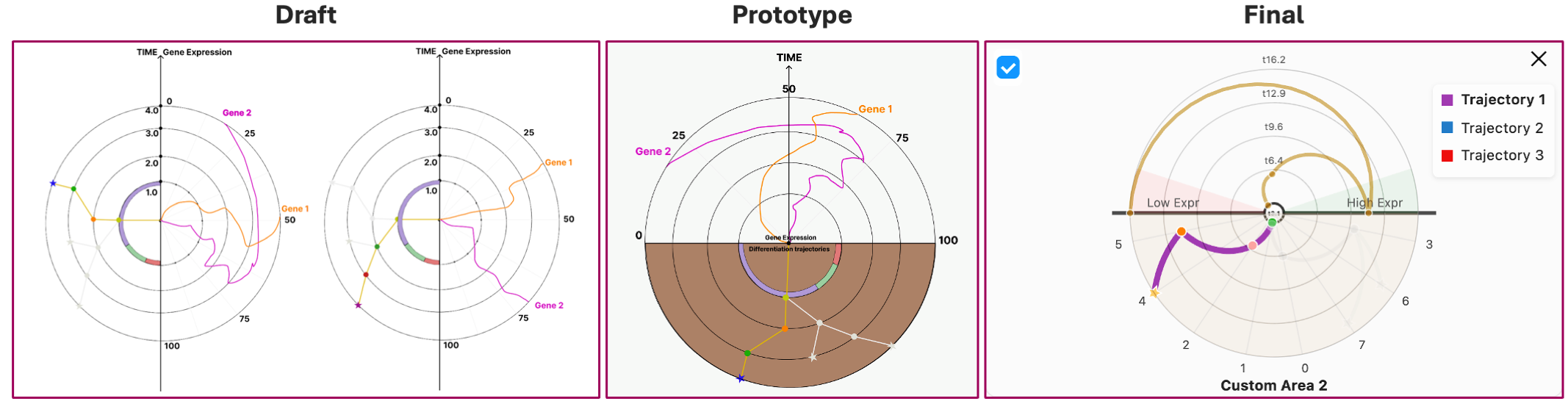}
  \caption{
    \label{fig:Glyph}
    Glyph design iterations. Initially, the polar graph was split into left and right to show changes in gene expression and cell state; the prototype rotated the graph 90° to invoke a plant root-to-stem metaphor; \textcolor{black}{the final version integrates neighborhood analysis into the lower semicircle, encoding the neighborhood intensity of each cluster within the selected region.}
    }
\end{figure*}

ST tools typically profile the gene expression state of each cell at a single point in time. In contrast, biological processes, such as differentiation, activation, repair, inflammation, and tumor progression, are dynamic and continuously evolving. While designing an effective visualization for pseudo-temporal analysis results is highly desirable, it is also inherently challenging. First, pseudo-time trajectories do not form a single linear path, but rather a forest-like structure with multiple root nodes, branching points, and distinct paths. Second, it is essential to represent changes in key genes along each trajectory over pseudo-time. Finally, the system must support a clear comparison of multiple pseudo-time results.

Through a parallel prototyping process followed by serial refinement guided by expert feedback related to clarity, interpretability, and overall legibility, we designed a novel glyph. In the early design stage, we explored several alternative representations, including linear timelines, radial tree diagrams, and layered sankey-style representations that directly reflected the branching topology of inferred pseudo-time paths. Although these designs preserved the original computational structure more explicitly, they also occupied irregular visual space and made repeated side-by-side comparison across regions difficult. This limitation led us to prioritize a more compact and regularized visual form that could support comparison without discarding the key temporal structure. We therefore moved away from an explicit branching representation and adopted a glyph design inspired by natural plant growth, where stems extend upward over time and roots spread downward. This metaphor provides a coherent visual logic for dividing the glyph into two coordinated halves, enabling multiple aspects of pseudo-temporal analysis to be integrated into a balanced circular layout (Fig.~\ref{fig:Glyph}) with the radial dimension encoding temporal progression. Specifically, in the lower half, each path corresponds to a selected region; star-shaped markers denote terminal cell states, and colored circular markers indicate intermediate stages. To further reflect cell–microenvironment interactions during state transitions, it is subdivided according to clusters imported from the UMAP Panel and arranged counter-clockwise by neighborhood intensity. Complementing this trajectory-focused representation, the upper half serves as a compact expression summary, with the left and right sides indicating low and high expression levels. \textcolor{black}{Since the biologists were particularly interested in identifying genes that are up- or downregulated, we added low- and high-expression range indicators to the left and right sides of the upper portion of the glyph to display the mean expression for each gene.} Additionally, hovering reveals a dashed line indicating the average expression level of the selected gene at that position, providing immediate quantitative feedback. Together, these design elements address several challenges in visualizing pseudo-temporal analysis (T2.1, T2.4, T2.5). 

Arriving at this design required resolving several competing considerations. One of the main difficulties in this design process was balancing three competing goals: preserving the branching nature of pseudo-time trajectories, showing associated gene dynamics, and keeping the visualization compact enough for repeated comparison across regions and samples. Emphasizing any one of these goals too strongly often weakened the others. For example, topology-preserving layouts improved faithfulness to the inferred trajectory structure, but made cross-region comparison harder; conversely, highly regular layouts supported comparison but risked obscuring branching details. Another challenge was integrating trajectory structure with neighborhood context imported from the UMAP-based analysis. Because these two data components follow different semantics, one describing temporal progression and the other describing local cellular environment, combining them in a single glyph without overwhelming the viewer required several rounds of simplification and expert-guided adjustment. The final design (Fig.~\ref{fig:Glyph}  right) reflects this iterative negotiation between expressiveness, interpretability, and comparability.

While the natural plant growth glyph was designed for ST pseudo-temporal analysis, it is generalizable to other temporal cluster modeling problems. The Supplemental Materials show applications to two other problems: airplane contrail modeling, respectively the impact of the El Niño and La Niña events on the Arctic glacier extent.

\section{Validation and Results}
Our next contribution is an evaluation of Loom, with the following components: (i) a demonstration of capabilities on two case studies developed by our three expert collaborators, who are senior researchers specializing respectively in ST technology  (co-author OK), cancer biology and wound-healing research (co-author AS), and in translational medicine with a focus on ST applications (co-author HC);  (ii) a performance evaluation; and (iii) a quantitative evaluation with an external group. The Supplemental Materials provide additional details for the  performance evaluation, and qualitative feedback from the domain experts. The case studies were developed through multiple online/remote and in-person sessions with the domain experts over the course of one year. The three expert co-authors provided feedback during the design process, while the external group was not involved in the design. 

\subsection{Case Study 1: Human skin epidermal analysis}
This case study focuses on evaluating a sample of human normal skin tissue, encompassing 114,213 cells and 17,079 genes, at a 2µm resolution. Our collaborators were particularly interested in analyzing keratinocyte markers, starting with two known keratinocytes which play roles in epidermis differentiation: TOP2A, which identifies the highly proliferative basal compartment~\cite{yang2025top2a}, and KRT10, a spinous layer–specific differentiation marker~\cite{LEUBE2016569}. Collaborators OK and AS conducted sample evaluation, ST assay and data processing, of the normal skin tissue, while HC provided additional interpretation of the ST analysis. This case study demonstrates reasoning across three layers: genes, cells, and local neighborhoods (microenvironments).

The analysis started with the local neighborhoods for a region of interest in the epidermal region (Fig.~\ref{fig:Teaser}.A). After delineating the region, the experts performed clustering analysis to assess whether the computationally inferred epidermal subpopulations recapitulate an established layer-wise organization. Specifically, 
epidermal differentiation follows a well characterized vertical progression, in which keratinocytes originate in the basal layer, undergo gradual maturation as they move upward through the spinous and granular layers, and ultimately form the stratum corneum~\cite{fuchs1990epidermal
}. This spatially organized process establishes a clear correspondence between cellular state and physical position within the tissue, making it a suitable biological reference for validating computational results. Loom showed that the resulting clusters exhibited a clear and orderly stratification, mirroring the known architecture of the epidermis (Fig.~\ref{fig:Teaser}.C).

\textcolor{black}{The clustering resolution for this ROI was not specified a priori. With a clustering K fixed at 15, HC iteratively adjusted the Leiden resolution parameter in the UMAP Panel and compared the resulting partitions to examine their spatial localization, GO enrichment, and consistency with known epidermal organization. During this process, HC noted} a small but distinct cluster localized within the basal layer (Fig.~\ref{fig:Teaser}.C) adjacent to but not overlapping with the basal cells, indicating that they share the same tissue microenvironment. Inspection of the Tissue Sample Panel confirmed these cells were scattered throughout the basal layer (T1.1, T1.2). This caught his attention because this type of cluster is not easily detectable in ST data. HC used Loom to perform GO spatial neighborhood enrichment in the microenvironment. The result identified the biological functions associated with the set of genes and provided a compelling explanation through our system: over-expressed melanin creation genes of TYR, TYRP1~\cite{murisier2006genetics, kobayashi1998tyrosinase} in a cluster showing strong enrichment for melanogenesis-related processes, indicating that it represented melanocytes (T1.3). HC was particularly pleased to note that melanocyte expression was revealed, despite dominant signals from the surrounding keratinocytes in the microenvironment.

Encouraged by this discovery, the group moved beyond static spatial inspection and initiated a spatiotemporal analysis to jointly examine how cellular differentiation unfolds along both inferred pseudo-time and physical tissue space. Using the TOP2A and KRT10 markers as anchors, they observed a clear bottom-to-top differentiation trajectory, with gene expression changes matching known biological patterns: TOP2A was expressed earlier than KRT10, and both genes showed an initial increase followed by a decrease (Fig.~\ref{fig:Teaser}.D). 

The group then focused on the spatial trajectory plot, using the differentiation direction and selected region as input (Fig.~\ref{fig:Teaser}.A). In the selected region, which contained a mixture of epidermal and dermal tissues with minor dermal intrusions within the epidermis, they checked whether the expression gradients of the key genes aligned with the pseudo-time sequence (Fig.~\ref{fig:Teaser}.B). Consistency between the two analyses would strengthen confidence in the inferred pseudo-temporal trajectories (T1.4, T1.5). In this region, characteristic basal-layer genes were expected to fluctuate along the tissue direction and gradually decrease near the horny outer layer of the skin, stratum corneum. The Loom spatial trajectory analysis identified KRT14—a known basal-layer marker~\cite{sumer2019keratin}—as one of the genes whose expression profile most strongly aligned with the inferred trajectory. Experts confirmed that its spatial fluctuations followed the expected basal to suprabasal gradient. Subsequent pseudo-temporal analysis further validated this pattern: in the upper part of the glyph, KRT14 oscillated multiple times between the left and right sides, mirroring the results observed in the spatial trajectory analysis (Fig.~\ref{fig:Teaser}.B,D) (T2.5). 

Notably, a pronounced inflection point appeared in the lower half of the glyph (Fig.~\ref{fig:Teaser}.D). By examining the two clusters that contributed to this inflection point (Fig.~\ref{fig:Teaser}.\textcolor{black}{A,C}), the group noted that their spatial distributions substantially overlapped, suggesting they likely represented the same biological population rather than two distinct clusters. This was considered a meaningful observation, as the presence of an inflection point can serve as a diagnostic signal for whether cluster boundaries require refinement. The finding, therefore, reflects genuine tissue biology rather than a pseudo-temporal modeling artifact.

\subsection{Case Study 2: Human thigh wound healing roadmap}

\begin{figure*}[t]
  \centering
  \includegraphics[width=0.8\linewidth]{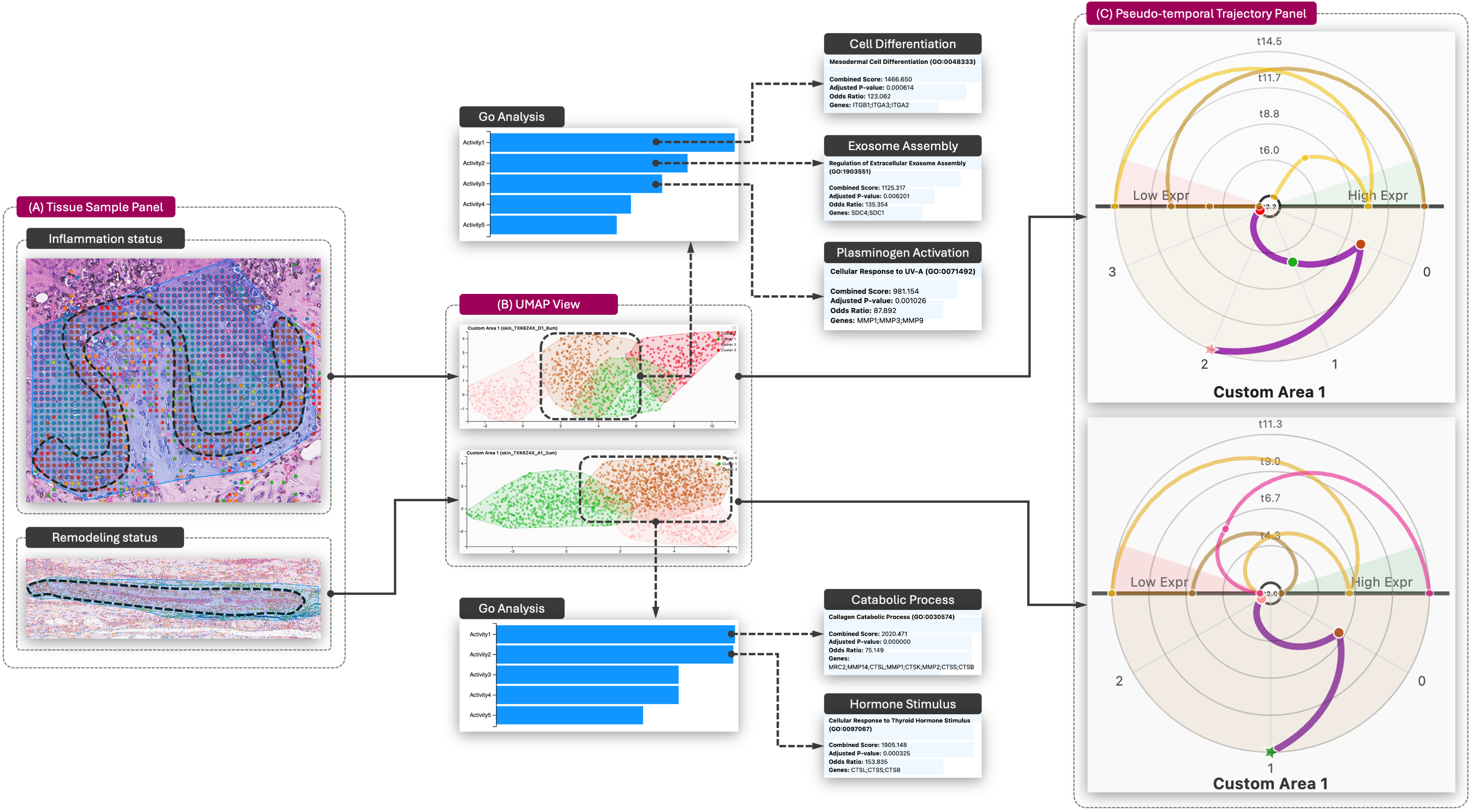}
  \caption{
    \label{fig:Case_Study_2}
    Inflammation ROI analysis Case 2. (A) Top: Inflammation Status; Highlighting principal repair-related cell populations (brown). Bottom: Remodeling Status. Showing a vascular ROI (green) surrounded by a blue-stained area. (B) Top: Inflammation Status; The trajectory (clusters 3 $\rightarrow$ 1 $\rightarrow$ 0 $\rightarrow$ 2) aligns with spatial progression. Bottom: Remodeling Status;  Three clusters (\textcolor{black}{2 $\rightarrow$ 0 $\rightarrow$ 1}) define a distinct temporal progression pattern. GO analysis of selected cell populations: Inflammation Status highlights processes related to cell differentiation, tissue integrity, and microenvironment regulation. Remodeling Status is enriched for collagen-related processes, including collagen fibril organization and catabolic activity.
    (C) Pseudotime analysis. Top: Inflammation Status; COL3A1 (\textcolor{black}{Ochre}) progressively increases toward the wound, while FOSL1 (\textcolor{black}{Amber}) peaks at the wound edge and drops at the center. Bottom: Remodeling Status; PECAM1 (\textcolor{black}{Ochre}) is highly expressed early and declines, ACTA2 (\textcolor{black}{Amber}) peaks at mid pseudo-time, and COL1A1 (pink) increases toward the late stage.
    }
\end{figure*}

In the second study, our collaborators examined biological processes occurring at human tissue wound sites. The skin wound healing process involves three main stages: (1) inflammation,  where blood clots form and immune cells like macrophages clean the wound and fight infection, (2) proliferation, where new blood vessels grow and granulation tissue fills the wound, and (3) remodeling, where tissue strengthens, reorganizes, and forms a scar. The experts analyzed two datasets: the first included 4116 cells and 18085 genes (8µm resolution), and the second contained 114213 cells and 17079 genes (2µm resolution). The group focused on regions surrounding skin injuries, examining immune-epidermal cell interaction during inflammation and stromal cell recombination during remodeling. Rather than validating outputs, the experts aimed to uncover functional patterns and cellular behaviors related to tissue repair and regeneration at both resolution scales. 

The experts first identified a distinct cell group in the first dataset using clustering and GO analysis (Fig.~\ref{fig:Case_Study_2}.A.Top). The Loom-executed GO enrichment indicated several wound-related immune activity and tissue rebuilding, suggesting that these cells were reacting to the injury, helping trigger the body's defense response and contributing to repair (Fig.~\ref{fig:Case_Study_2}.B.Top) (T1.1, T1.2). With this insight, the experts then used the trajectory glyph to examine how these wound-related processes evolve jointly across spatial context and inferred temporal progression and observed typical gene changes that occur during the healing process (Fig.~\ref{fig:Case_Study_2}.C.Top) (T1.3, T1.4). Next, they investigated gene FOSL1, previously reported to help skin close up again, and confirmed that it was strongly expressed near the wound edge. However, at the wound surface, the gene was no longer active. Along the pseudotime trajectory, FOSL1 expression rose well above the high-expression (green) range at the wound edge, but dropped back to the low-expression (red) range inside the wound region, revealing a coordinated spatiotemporal pattern in which peak expression is localized at the wound boundary and diminishes as cells transition into the rebuilt region. The experts noted a clear turning point in the trajectory shown beneath the glyph. Unlike typical differentiation processes, this shift did not arise from a cluster transition that required merging. Instead, they concluded that the wound region itself tends to be the most “extreme” area in the tissue, which naturally creates this sharp change in the trajectory (Fig.~\ref{fig:Case_Study_2}.C.Top)  (T1.5, T2.4, T2.5).

The experts then analyzed the second sample (Fig.~\ref{fig:Case_Study_2}.A.Bottom). Unlike the first dataset, the histology images revealed a distinct blue-stained area in the subcutaneous tissue, with nuclei unusually arranged in bands along blood vessel walls, suggesting a wound-associated region. Using the same Loom clustering and GO analysis workflow, the group found that this area indicated features of the remodeling stage, a transitional phase between late proliferation and tissue remodeling (Fig.~\ref{fig:Case_Study_2}.B.Bottom). To further validate this interpretation, the experts again combined spatial evidence with pseudo-temporal analysis to examine how these remodeling signals are organized across both dimensions. The experts noted a clear and interpretable chain of evidence supporting this conclusion. Cluster 2 showed angiogenesis along with markers of vascular maturation, indicating that excess capillaries were degenerating while others stabilized—a hallmark of remodeling. Cluster 0 exhibited strong muscle contraction signals consistent with myofibroblasts that pull wound edges together and reduce wound size. Cluster 1 corresponded to cells in the blue-banded area showing a distinctive "build-and-destroy" pattern, with high expression of both matrix-building and matrix-degrading genes, reflecting active tissue renovation (T2.1, T2.2, T2.3). These spatially localized functional signatures were then examined along inferred trajectories, which confirmed these hypotheses (Fig.~\ref{fig:Case_Study_2}.C.Bottom) (T2.5).

\subsection{Performance}
\textcolor{black}{
New Visium HD datasets can be prepared with reasonable computational cost and then analyzed interactively in Loom at the ROI level (Supp. 2\#1.3). Dataset preprocessing completes in approx. 2 minutes on average at 8 µm and 16 µm, while the finer 2 µm setting may require 35 min. Loom loading and rendering complete in 13.26~sec ± 2.33~sec. ROI-level analyses across regions containing 3,029–17,658 cells remain practical for interactive use, with the main downstream steps completing within seconds to approximately 2 minutes. Together, these results characterize the performance and scalability of Loom’s import and analysis workflow.
}

\subsection{Usability study results}
To assess the usability of Loom and its novel glyph beyond our co-designer collaborators, we collected feedback from an external group  using the System Usability Scale (SUS). Seven participants with varying levels of experience in biological data visualization and analysis participated in the study. \textcolor{black} {The study was conducted following UIC IRB regulations, and informed consent was obtained.} The participant number was necessarily limited by the availability of specialized knowledge of ST, clustering, and pseudo-temporal analysis. Participants were first introduced to the visual front end and then asked to complete four analytical tasks. Participants rated ten standard SUS items on a scale from 1 to 5. Additionally, we included open-ended questions to obtain feedback on specific features and potential improvements related to spatiotemporal analysis tasks.

The system achieved a SUS score of 80.9, characterizing its usability as excellent. Quantitative feedback indicated that the external group felt confident using the system (M = 4.38 ± 0.52) and praised its functional integration (M = 4.63 ± 0.52). Participants reported that they would be likely to use the system when studying genes’ spatiotemporal changes (M = 4.75 ± 0.46) and agreed that the interface was easy to use (M = 4.50 ± 0.76), as well as that most users would be able to learn the tool quickly (M = 4.13 ± 0.99). Conversely, participants generally disagreed that the system was unnecessarily complex (M = 1.50 ± 0.53), inconsistent (M = 1.50 ± 0.53), or cumbersome (M = 1.50 ± 0.53). They also reported only a moderate need for support from the developer (M = 2.63 ± 0.92) and a moderate initial learning burden before getting started (M = 2.88 ± 1.25). Additionally, participants expressed a positive overall impression that the system met their expectations (M = 4.38 ± 0.92).

The system received positive feedback for its interactive and visually grounded design, which effectively supports region-based spatiotemporal analysis. Participants particularly appreciated the ability to define regions of interest and examine directional gene expression changes within those regions, enabling detailed investigation of biological processes such as tissue repair and remodeling. Overall, the evaluation demonstrates strong usability and user acceptance, with improvements aimed at enhancing learnability rather than core interaction design.

\section{Discussion and Conclusion}
To understand what is “happening” within a ST sample, experts must reason across three progressive layers: genes, cells, and the microenvironment, which is a complex process. Loom enables investigators to focus on local neighborhoods or regions of interest and directly obtain interpretable insights, while also allowing them to establish global links and trajectories across regions and samples in a spatiotemporal analytical context. Loom supports this type of analysis through a combination of complementary views, a compact novel encoding, and a flexible computational backbone.

Although existing visualization tools focus on specific tissue regions~\cite{warchol2022visinity, somarakis2019imacyte}, they do not support analysis of the fine-grained directional variations within these regions. In contrast, Loom integrates ROI-based exploration through the Tissue Sample Panel with spatial trajectory analysis. Through the Spatial Trajectory Panel, experts can freely specify directional paths within an ROI based on their domain knowledge and observe the most pronounced gene expression changes along those directions. This provides a convenient way for domain specialists whose analytical questions often center on specific biological processes, such as the direction of tumor invasion, to examine spatially organized transitions. However, high-resolution spatial data introduce a fundamental challenge: performing trajectory inference directly at such scales generates substantial noise, suppresses feature-gene detection, and increases computational cost. To address this issue, we developed a spatial coordinate registration algorithm that dynamically projects user-defined high-resolution geometric inputs onto a lower-resolution computational space. This design preserves computational efficiency while maintaining accurate and interpretable trajectory inference, as validated by domain experts. 

Our results show that integrating pseudo-temporal simulation with spatial context yields substantially deep biological insights. Conventional clustering often fails to explain why certain cell groups aggregate. To address this gap, we introduced a clustering ranking strategy based on neighborhood enrichment, enabling us to quantify the “spatial sociality” of cell populations. This reframes the results as interacting neighborhoods rather than isolated cell states. Across case studies, the top-ranked clusters identified by Loom capture key interaction hubs within tissues: clusters whose constituent cells play central roles in maintaining tissue homeostasis. These findings \textcolor{black} {suggest} that Loom can reveal complex biological variation, particularly the links between cellular developmental states and their local microenvironments, which are frequently overlooked by standard dimensionality-reduction techniques such as UMAP or t-SNE. \textcolor{black}{Our expert-validated findings are, however, limited to two human tissue cases.}

To communicate these spatiotemporal patterns, we designed a \textcolor{black}{compact glyph that provides an expressive visual representation of the data’s multidimensional structure. The glyph's small} visual footprint allows it to scale effectively as the analysis area expands. Although the encoding is designed for a particular biological context, the glyph is generally applicable. It can be extended to other domains involving time-series patterns with multiscale variation, such as in urban mobility analytics. \textcolor{black}{Following a brief glyph introduction, external participants with no prior ST experience were able to interpret the glyph, identify patterns, and complete the analytical tasks, even when they lacked prior biodata experience (Supp. 2\#4).} In terms of scalability, the Visium HD datasets used in our system are representative of contemporary ST analysis problems, particularly those involving high-resolution tissue profiling. Compared with conventional ST data, they contain substantially more spatial measurement units and much denser molecular signals, indicating Loom can handle scalable analysis.

Loom leverages several well-established and maintained external tools and packages for analysis. Pseudo-temporal results may appear reversed because the embedded Slingshot algorithm \textcolor{black}{does not intrinsically determine the direction of a trajectory: pseudotime 0 is assigned to a heuristically designated cluster, which may not correspond to the true biological origin and thus can result in a reversed pseudotime direction.} We address this issue by enabling manual start-cluster selection, supported by integrated GO enrichment to help users identify biologically plausible roots. Second, spatial trajectory analysis via SPATA2 leverages our system's Spatial Coordinate Registration Algorithm to ensure computational stability and efficiency. SPATA2 still mandates a minimum cell threshold within the region to guarantee reliable results. Loom uses basic functionality from these tools, meaning tool updates usually involve simple changes to function parameters. Our lab has the resources to support and maintain Loom in the coming years.

In conclusion, we designed, implemented and evaluated Loom, a visual computing system designed to address the challenges of multi-sample, multi-scale ST analysis. By integrating a computational backbone and a registration algorithm with coordinated multi-view visualizations, Loom enables users to move fluidly from global tissue comparisons to fine-grained examinations of local microenvironments. The natural plant-growth-inspired glyph serves as a unifying representation that links pseudo-temporal patterns, spatial enrichment, and gene expression dynamics within user-defined regions of interest. Through collaborations with pathology and oncology experts and external SUS evaluation, we demonstrate that Loom supports meaningful biological discovery across samples and conditions, including the identification of cellular transitions and scale-specific spatiotemporal expression patterns. Together, these contributions offer a practical, extensible approach for interpreting complex ST datasets and advancing biological inference in spatially resolved studies.

\acknowledgments{
Our work is supported by NIH NCI R01CA258827, NIH UG3 TR004501, NSF CNS-2320261, and the UIC Institute for Health Data Science Research. 
}

\bibliographystyle{abbrv-doi-hyperref}

\bibliography{template}

\end{document}